\documentclass[onecolumn,epsfig,pre]{revtex4-2}
\usepackage{graphicx}
\usepackage{amssymb}
\usepackage{amsmath}
\usepackage{hyperref}
\usepackage{latexsym}
\usepackage{mathtools}
\usepackage{mhchem}

\def\be{\begin{equation}}
\def\ee{\end{equation}}
\def\bea{\begin{eqnarray}}
\def\eea{\end{eqnarray}}

\begin{document}

\title {Clustering and finite  size effects  in a two-species exclusion process}

\author{Jim Chacko}
\affiliation{Christian College, Angadical South PO, Chengannur, Kerala 689122, India.}

\author{Sudipto Muhuri}
\email{sudipto@physics.unipune.ac.in}
\affiliation{Department of Physics, Savitribai Phule Pune University, Pune, India}

\author{Goutam Tripathy}
\affiliation{Institute Of Physics, Sachivalaya Marg, Sainik School PO, Bhubaneswar 751005, India.}
\affiliation{ Homi Bhaba National Institute, Anushaktigar, Mumbai 400094, India }


\begin{abstract}
We study the cluster size distribution of particles for a two-species exclusion process which involves totally asymmetric transport process of two oppositely directed species with stochastic directional switching of the species on a 1D lattice. As a function of $Q$ - the ratio of the translation rate and directional switching rate of particles, in the limit of $Q \rightarrow 0$, the probability distribution of the cluster size is an exponentially decaying function of cluster size $m$ and is exactly similar to the cluster size distribution of a TASEP. For $Q>>1$, the model can be mapped to persistent exclusion process (PEP)  and the average cluster size, $\langle  m  \rangle \propto Q^{1/2} $. We obtain an approximate expression for the average cluster size in this limit. For finite system size system of $L$ lattice sites, for a particle number density $\rho$, the probability distribution of cluster sizes exhibits a distinct peak which corresponds to the formation of a single cluster of size $m_s = \rho L$. However this peak vanishes in the thermodynamic limit $ L \rightarrow \infty$. Interestingly, the probability of this largest size cluster, $P(m_s)$, exhibits scaling behaviour such that in terms of scaled variable $Q_1 \equiv Q/L^2 \rho(1-\rho)$, data collapse is observed for the probability of this cluster. The statistical features related to clustering observed for this minimal model may also be relevant for understanding clustering characteristics in {\it active } particles systems in  confined 1D geometry. 

\end{abstract}

%\keywords{Cellular and subcellular biophysics}
\maketitle
%\pacs{87.15.A-}{Cellular and subcellular biophysics}
%\pacs{87.16.A-}{Theory, modeling, and simulations }
%\pacs{87.16.Ka}{Filaments, microtubules, their networks, and supramolecular assemblies}
%\pacs{87.17.Ee}{Cell division}

%\maketitle
\section{Introduction}
Driven lattice gas models have been the subject of considerable attention due their remarkably rich stationary and dynamical behaviour \cite{intro1, intro2, intro3, intro4, intro5, intro6, intro7}. These models have also served to illuminate generic coarse grained features of many biological processes ranging from bidirectional transport of motors and cellular cargoes on cytoskeletal filaments \cite{bio1, bio2,bio3, bio4,bio5,madan-pre} to growth of fungal hypha \cite{fungi1,fungi2,fungi3}, and dynamics of bacterial suspensions \cite{soto}.
For instance, motivated by the phenomenon of motor driven bidirectional transport of cellular cargoes on cellular filaments \cite{ref_welte,ref_hollenbeck,tug1,tug4},  a minimal model focusing on the  interplay of the switching dynamics of individual cargo  and their motility on discrete 1D lattice has been studied theoretically using a two-species driven lattice gas model \cite{madan-pre}.  The steady state density and current profile for this model has been characterized in terms parameter $Q$ which is defined as the ratio of the translation rate of particles on the lattice $(v)$ and directional switching rates $(\alpha)$. It was observed that in the limit of very slow switching rates (compared to the translation rates), the system approaches a jammed phase with a net current $J \sim 1/Q$ \cite{madan-pre}. This naturally invites the question on how $Q$ would affect the cluster size distribution of particles on the lattice. We specifically focus on the generic statistical features associated with  the cluster size distribution within the framework of a model which is very similar to a two species model proposed in Ref. \cite{madan-pre} in context of bidirectional transport of cellular cargo on biofilaments and in Ref.\cite{soto} in context of modeling {\it run-and-tumble} dynamics of motile bacterial system. However unlike in the case of these models, we consider a situation where the directional switching of particles on the 1D lattice occurs {\it only if} the particles are in the vicinity of each of other, i.e., they are part of a cluster. Indeed for experiments performed with self-propelled water droplets, confined in a 1D microfluidic channel, the swimming droplets are seen to reverse their directionality  of movement only upon suffering collisions with other droplets in the channel \cite{droplet}. In context of collective cell migration in 1D collison assays such as that  MDCK (Madin-Darby canine kidney) cells, it has been observed that  the cells can switch their directionality of movement only upon interaction with other cells \cite{ananyo, CIL}. Also, constraints on experimental feasibility restricted the maximum size of the collision assays thereby also making the effect of the finite size of the system relevant for determining the transport and clustering characteristics of the cells in these experimental geometries \cite{ananyo}. These experimental examples serve as the broad motivation for the specific dynamic rules of switching of particles that we incorporate in our minimal model and for undertaking investigation of the finite system size effects on the cluster characteristics for the model. 

In the next section we first describe the model details, delineating the difference of the model that we study with the one studied in Ref.\cite{madan-pre} and Ref.\cite{soto}. We then proceed to  obtain  the cluster size distribution of particles on the 1D lattice using Monte Carlo simulations. We show that for $Q >>1$, the cluster characteristics would be similar to the model proposed in Ref.\cite{soto}. In fact in this limit, following the prescription proposed in Ref.\cite{soto}, we can obtain an approximate analytical expression for average cluster size. Subsequently, we look at the finite system size effects on the cluster size distribution and uncover an interesting scaling behaviour exhibited by the largest size cluster.  Finally, we conclude our discussion with summary of our results.  

\section{Model and Methods}
\subsection{Two species Bidirectional Exclusion Process (2S-BEP)}

\begin{figure}[t!]
    \centering
    \includegraphics[width=7cm, height =3cm]{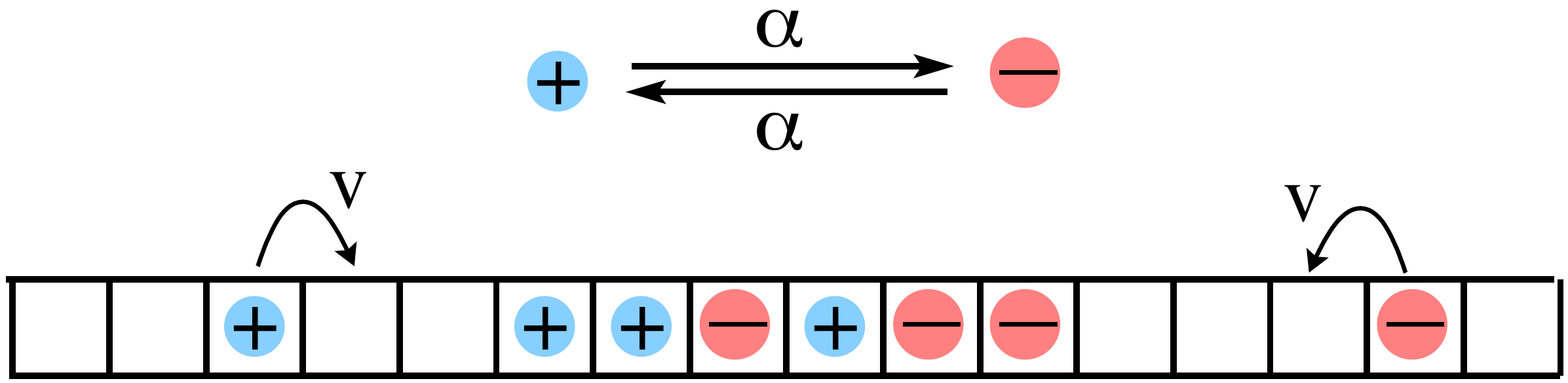}
    \caption{Schematic representation of the dynamical processes of switching and translation for the Two Species Bidirectional Exclusion Process (2S-BEP).}
    \label{Fig1}
\end{figure}    

We consider  a discrete 1D lattice of $L$ sites with $N$ particles.  The particle can either be in the $\oplus$ state (moving to right) or in the $\ominus$ state (moving to left). These  states of the particles which are defined in terms of their directionality of movement on the lattice, are regarded as the two species of particles. The interaction between the particles is described in terms of hard core exclusion, thus restricting the maximum occupancy of any lattice site to 1. Thus each lattice site $i$ is either vacant or can be occupied by a  $\oplus$ species or a  $\ominus$ species. While the $\oplus$  can translate to the adjacent vacant site on the right or with a rate $v$,  $\ominus$  can translate to the adjacent vacant site on the left or with  rate $v$. Whenever two particles are on adjacent sites of the lattice, with a fixed rate of switching $\alpha$ they can switch to other state, i.e., particle in $\oplus$ state can switch to $\ominus$ state and vice-versa. Its worthwhile to point out that the switching of particle can occur {\it only if} the adjacent site of the particle is occupied by another particle unlike the dynamics in PEP \cite{soto} or the dyanmics considered in Ref. \cite{madan-pre}. For PEP, the polarity for an individual particle ( which are not part of any cluster) can switch with finite probability unlike in our case. However it may be surmised that in the limit of low switching rate, the dynamics would be very similar to our case. Another point of contrast is that while we have been performed the simulation dynamics for the model by considering random update of particles, Ref.\cite{soto} considers, a sequential update of particles.  We consider periodic boundary condition for the lattice. The various dynamical processes are schematically displayed in Fig.\ref{Fig1}. We note that the overall total density of particles ( $\oplus$ and $\ominus$)$, \rho$ is conserved under the dynamics although the individual number density of each species of particle is not conserved due to the stochastic directional switching process. 

\subsection{Simulation Details}
We initially start with a random distribution of particles on the lattice with the specified number density of particles. A site on the lattice is selected at random. If the chosen site is occupied by a particle then we implement the monte carlo (MC) move for a particular process ( translation of directional switching) proportional to its rate.  We wait for an initial transient of $1000 \frac{L}{r}$ swaps , where $r$ stands for the lowest rate ( among switching and translation rates), to ensure that the systems reaches steady state. Subsequently we gather statistics for the cluster sizes, time averaging typically over atleast  5000 samples. These samples are collected with a time spacing of $10 \frac{L}{r}$. 

\begin{figure}[t!]
    \centering
    \includegraphics[height = 8 cm , width = \linewidth]{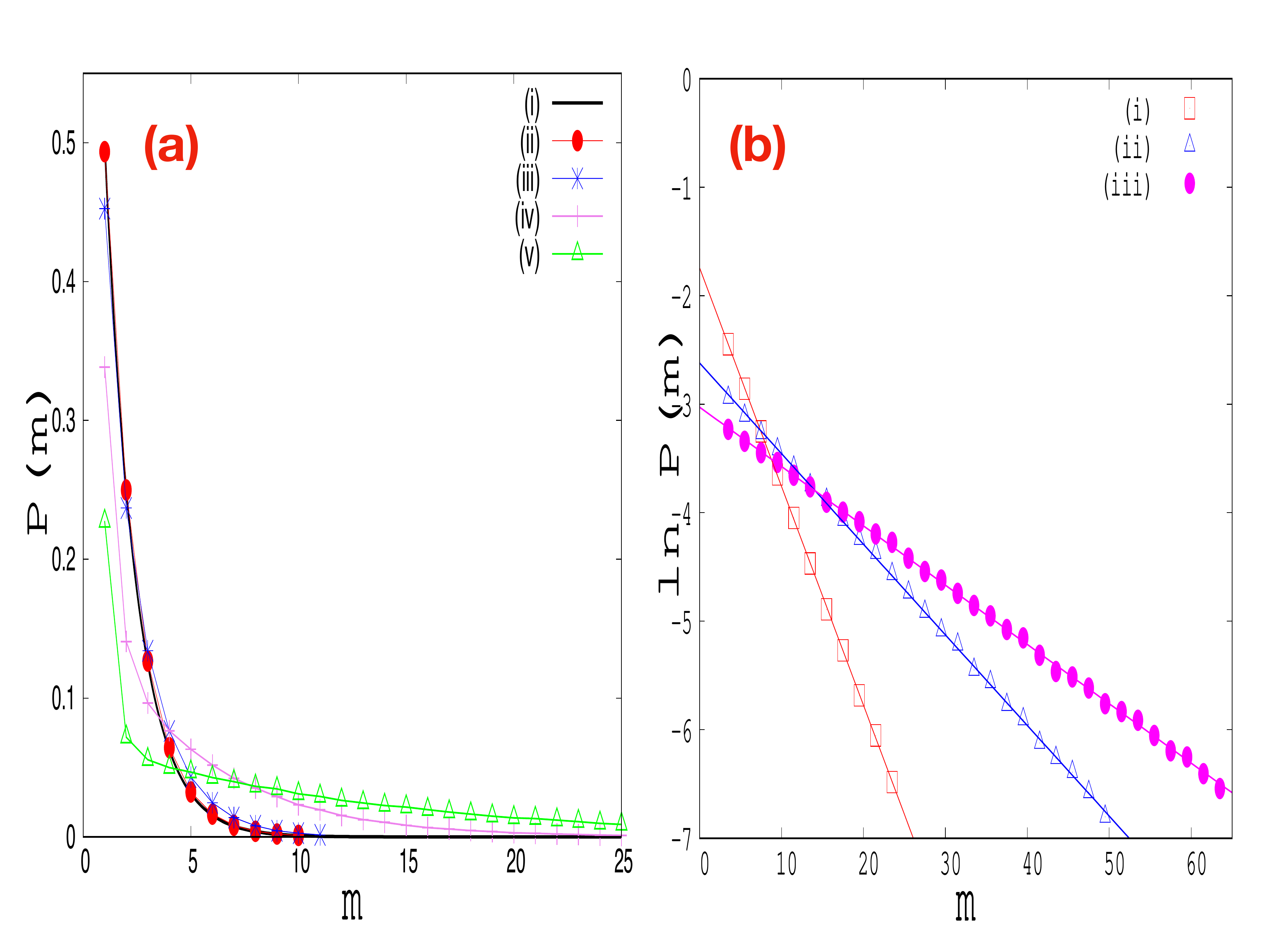}
    \caption{(a) Probability distribution of cluster size: (i)  corresponds to the exponential form of P(m) in Eqn.(1). Plots of cluster size distribution for (ii) Q = 0.1, (iii) Q = 1,  (iv) Q = 16 and (v) Q = 100 are obtained MC simulations. (b) logplot of P(m): (i) Q = 16, (ii) Q = 100, (iii) Q = 225. Here $\rho = 0.5$. MC simulations where done with $L = 1000$ and averaging was done over 5000 samples.}
    \label{Fig2}
\end{figure}    

\begin{figure}[t!]
    \centering
    \includegraphics[height = 8 cm, width = 10 cm]{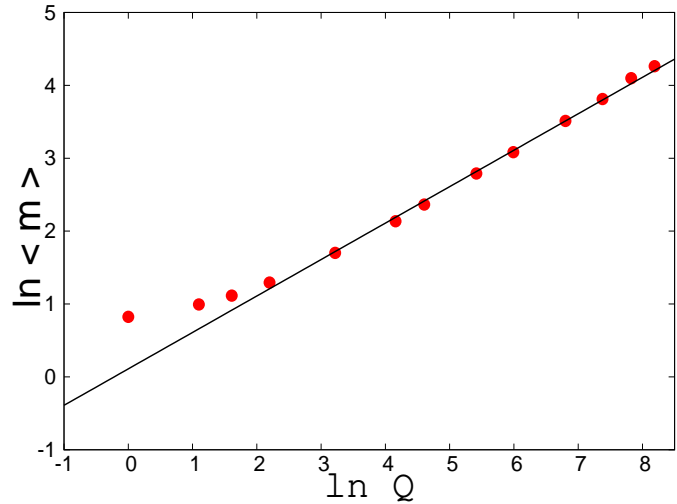}
    \caption{Plot of average cluster size $\langle m \rangle $ as function of $Q$: The dots corresponds to the data points obtained for different values of Q while the straight line corresponds to a line with slope 1/2.   Here $\rho = 0.5$ and MC simulations where done with $L = 2000$ and averaging was done over 5000 samples}
    \label{Fig3}
\end{figure}    
    
\section{Results}

In general, the inherent translational dynamics of oppositely directed particles lead to formation of clusters - comprising of a group of particle(s) bounded by two vacant sites. Further for a cluster comprising of two or more particles the  only way by which it can disintegrate is when particles at either end of the cluster hop out of the clusters. Such an event can occur only if the polarity of the particles at the cluster ends is pointed outwards which in turn can lead to movement of these particles out of the clusters and thereby leading to disintegration of the cluster. On the other hand, the size of the cluster can increase due to the inflow of (mobile) particles. Qualitatively the effect of increasing $Q$ corresponds to increasing the translation rate with respect to switching rate and this manifests in a tendency to form larger clusters in the lattice. When $Q >>1$, the system gets partitioned into a dense(d) phase comprising of particles and low density gas(g)  phase. We first present the results for cluster size distribution in the limit of $L \rightarrow \infty$.  
  
The limit of very high switching rate $\alpha$ compared to translation rate $v$ corresponds to the situation for which $Q \rightarrow 0$. In this limit, for $L\rightarrow \infty$, the probability of m-particle cluster is simply proportional to $\rho^m(1- \rho)$. With appropriate normalization, the normalized probability of cluster size $m$,  $P(m) = \rho^{m-1}(1-\rho)$ which maybe expressed in the exponential form, 
\begin{equation}
P(m) = \left(\frac{1-\rho}{\rho}\right)e^{-m/\xi},
\label{eq1}
\end{equation}
where $\xi = {| ~{ln \rho} ~|}^{-1}$. The  expression for the mean size of the cluster is $\langle m \rangle = {(1-\rho)}^{-1}$. In Fig.\ref{Fig2}a we show the plot of the exponential form of $P(m)$ as a function of $m$, obtained in Eq.\ref{eq1} along with the plots of $P(m)$ obtained by MC simulation for a range of different values of $Q$. As expected when, $Q$ is small (See Fig.\ref{Fig2}a), the size distribution of clusters tends to the limiting exponential distribution of  Eq.\ref{eq1} and is exactly similar to the cluster size distribution for TASEP \cite{pulki}. However with increase in the value  of $Q$, the probability distribution of the cluster sizes deviates from the limiting distribution of Eq.\ref{eq1}. Further increase of $Q$ leads to an increase in the average size of the cluster. In the large $Q$ limit, from the log-plot of the probability distribution $P(m)$, we can infer that the distribution continues to remain exponential in nature (See  Fig.\ref{Fig2}b). Next, we analyze the behaviour of the mean cluster size $\langle m \rangle$ on variation of Q. As far as the behaviour of mean cluster size is concerned,  we find that $\langle m \rangle  \sim Q^{1/2}$, for sufficiently large Q (see  \ref{Fig3}). In next subsection. (\ref{sub}), we present a derivation of the average cluster size based on the principle of maximization of configurational entropy which rationalizes this observation.

\subsection{Expression for average cluster size}
\label{sub}

For this system, when the directional switching rate of the particles in the lattice, $\alpha$ is much smaller than hopping rate of particles $( Q >> 1)$, the system gets partitioned into alternate regions of dense cluster phase $(d)$ which comprises of continuous stretch of particles (with no vacancies) and regions of very low density of particles - the gas $(g)$ phase. It may be noted that for the stationary state, at the location of domain wall separating gas region with dense cluster region, the flux of particle from gas region to the dense cluster region should be equal to the flux of particles from dense cluster region to the gas region. Equating the incoming and outgoing flux, we obtain the condition, $ \rho_g = 2/Q$. In this limit, the system gets mapped on to persistent exclusion process (PEP) discussed in Ref.\cite{soto} and the steady state is characterized by balance of incoming flux of particles from the gas region to the cluster region with the outgoing flux of particles from the cluster region of the gas region. Further, the clusters themselves evolve independently of the other cluster and their sizes are governed by the emission and absorption process of particles that occur at the dense cluster boundaries \cite{soto}. Consequently this process can be expressed in terms of  an equivalent equilibrium process for sizes of clusters wherein the cluster size distribution of the clusters is obtained by maximizing the configurational entropy for the system \cite{soto}. For the dense cluster phase (d), the configurational entropy $S$ corresponds to the number of ways in which $C$ different clusters can be arranged where clusters of same length are indistinguishable and are subject to the constraint that the total number of sites occupied by the dense cluster ($N_d)$ is held fixed and  the total number of clusters, $(C)$ is held fixed. Then it follows that, 

\vspace {0.3cm}

\begin{equation}
S = ln\left[ \frac{C !}{ \prod_{l} F_d(l)!} \right] - \lambda \left ( N_d - \sum_{l} l F_d(l) \right ) - \gamma \left ( C - \sum_{l} F_d(l) \right )
\end{equation}
Here $F_d(l)$  is the number of clusters of length $l$ in the dense cluster $(d)$ phase. 

Maximizing the the entropy by setting $\delta S = 0$ for independent variations of $\delta F_d(l)$ yields, $F_d(l) = A_d e ^{-l/l_d}$. Similarly for the gas (g) phase, 
$F_g(l) = A_g e ^{-l/l_g}$. The constants $l_d, l_g, A_d, A_g$ can determined by making use of the following considerations: (i) The total number of clusters in the gas phase must equal total number of clusters in the dense cluster phase and this  implies that  $\sum_{l} F_d(l) = \sum_{l} F_g(l)$. (ii) The total number sites occupied by clusters in the gas phase together with the total number sites occupied by clusters in the dense cluster phase  must equal total number of lattice sites. This would imply that $\sum_{l} l F_d(l) + \sum_{l} l F_g(l) = N$. (iii) The individual number densities in the gas region ($\rho_g$) and the cluster region  ($\rho_d$) should be consistent with the overall particle number density $\rho$. This is equivalent to the condition, $\langle l_d \rangle \rho_d + \langle l_g \rangle \rho_g = [ \langle l_d \rangle + \langle l_g \rangle ] \rho $. (iv) Balance of the production rate of dimers with the dissociation rate of dimer in the gas region: When $Q >>1$, the gas region has typically very low density of particles so that $\rho_g$ is small. When a switching event at the boundary of dense cluster occurs, there would be emission of particle into the gas region.  This will lead to production of dimer within the gas region which is enclosed between two dense cluster region. In steady state, the production rate of dimer must equal dimer dissociation rate. This is an approximation which would be valid in the low switching rate limit. The expression for the production of a dimer in the gas region $(W_p)$ is,

\begin{equation}
W_{p} = 2 \alpha^{2}\frac{ \langle l_g \rangle}{v} \sum_{l} F_g(l),
\label{dimer1}
\end{equation} 
while the overall disintegration rate of the dimer clusters is,  

\begin{equation}
W_{d} = 2 \alpha ~F_d(2)
\label{dimer2}
\end{equation}

Equating Eq.\ref{dimer1} and Eq.\ref{dimer2}, we obtain, 

\begin{equation}
 \langle l_g \rangle = \frac{ \sum_{l} F_g(l)}{Q F_d(2)}
\label{cond4}
\end{equation}

  \begin{figure*}[t]
    	\centering
    	\includegraphics[height = 8cm, width=\linewidth]{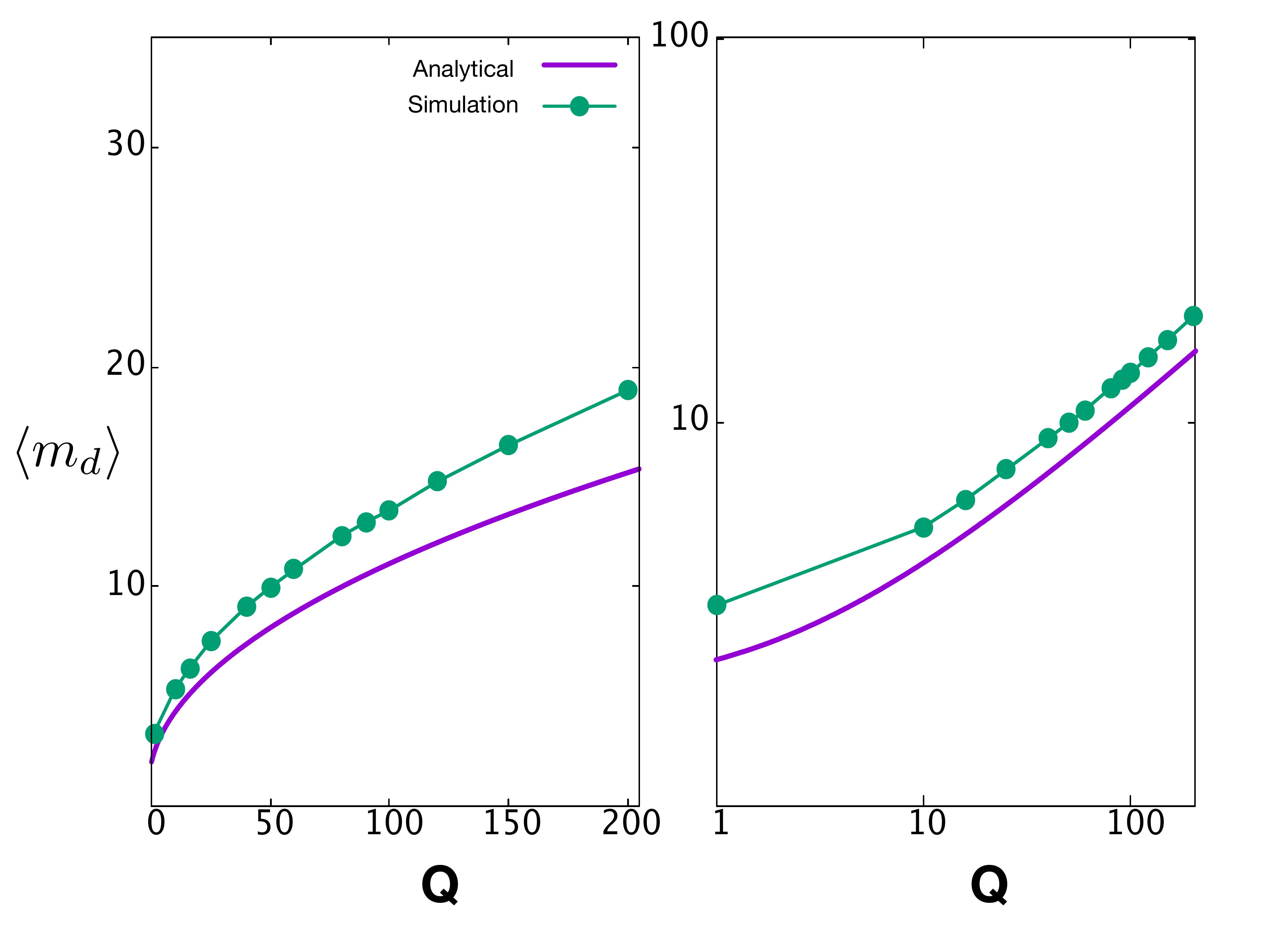}
    	\caption{Comparison of average cluster size in the dense cluster phase $\langle m_d \rangle $ as function of $Q$ obtained by MC simulations  with the approximate analytical expression in Eq.\ref{MF-cluster}. The corresponding log-log plot is also displayed. Here L = 2000 and $\rho = 0.5$. MC simulations were performed and averaging was done over 2000 samples }
    	\label{fig:new}
    \end{figure*}
       
In order to obtain an approximate expression for the mean cluster size,  we make use of the four conditions outlined above.  First we approximate the summations  by integrals to obtain a set of four algebraic equations involving $l_g$, $l_d$,
 $A_d$ and $A_g$ and which upon further simplification can be expressed as , 
\begin{eqnarray}
A_d l_d e^{-2/l_d}  &=& A_g l_g e^{-1/l_g} \nonumber \\ 
A_d l_d e^{-2/l_d} (2 + l_c) + A_g l_g e^{-1/l_g} (1 + l_c) &=& N   \nonumber \\
(2 + l_d) + 2(1 + l_g)/Q &=&  [(2 + l_d) + (1 + l_g)]\phi   \nonumber\\
\langle l_g \rangle &=& Q/l_d
\end{eqnarray}

 Solving these set of equations and assuming $Q >> 1$, we obtain the approximate expression for mean cluster size in the dense cluster region which reads as, 
 
\begin{equation}
\langle m_d \rangle  = 2 +  \sqrt{ Q \left( \frac{\rho}{1- \rho}\right)} 
\label{MF-cluster}
\end{equation}

Fig.\ref{fig:new} shows the comparison of the approximate analytic expression  mean cluster size in the dense cluster region with MC simulations.  Eq.\ref{MF-cluster} slightly underestimates the mean cluster size but is able to correctly predict the $\sqrt{Q}$ dependence of the mean cluster size.

%    \begin{figure}[t]
 %   	\centering
 %   	\includegraphics[width=14 cm, height = 8 cm]{fig7-new1.pdf}
 %   	\caption{ Probability of the largest size cluster $(m_s)$ as a function  of scaled variable $Q_s = Q/L^2 \rho ( 1 - \rho)$ : (i) L=100, $\rho = 0.5$, (ii) L=200, $\rho = 0.5$, (iii) L=400, $\rho = 0.5$, (iv) L=400, $\rho = 0.3$, (v) L=400, $\rho = 0.1$. Here the data points corresponds to the plots of Fig 6(a) and Fig6(b).  MC simulations were performed and averaging was done over 10000 samples.}
%    	\label{fig:npnm}
%    \end{figure}    

 \subsection{Finite size effect on cluster distribution}

\begin{figure}[t!]
    \centering
    \includegraphics[height = 8cm, width = 10cm]{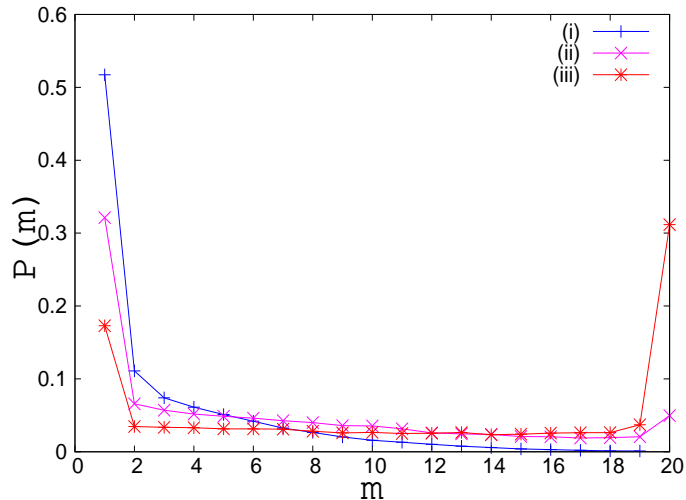}
    \caption{Probability distribution of cluster size P(m) as a function of cluster size, m: (i) For Q = 200. (ii) For Q = 500, P(m) starts exhibiting a secondary peak at maximum cluster size $m_s$ ( $m_s = \rho L = 20$). (iii) For Q = 2000, P(m) exhibits a sharp secondary peak at maximum cluster size. Here  L = 200 and  $\rho = 0.1$.  MC simulations were performed and averaging was done over 10000 samples.}
    \label{fig5}
\end{figure}    

    \begin{figure*}[t]
    	\centering
    	\includegraphics[width=\linewidth, height = 7 cm]{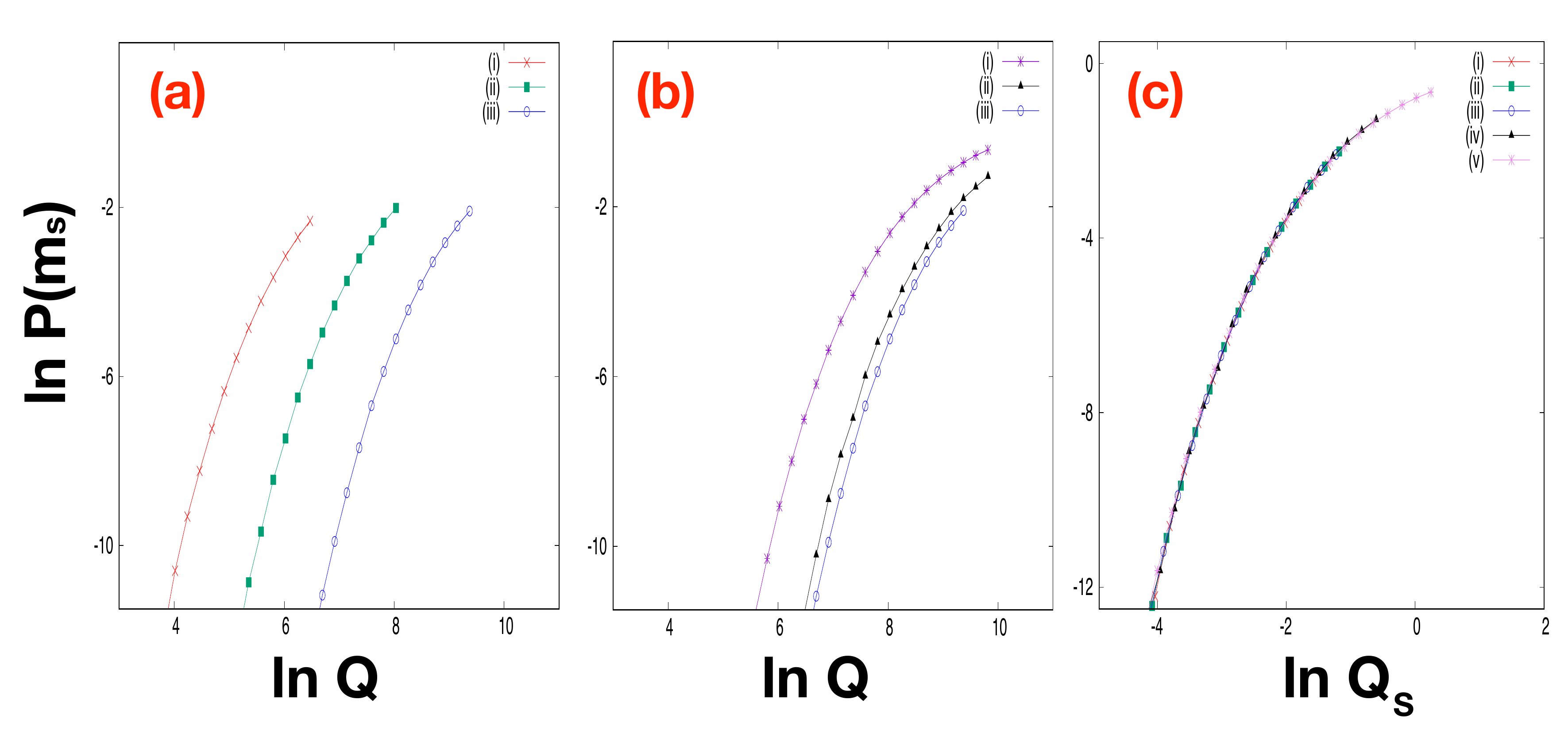}
    	\caption{ Probability of largest size cluster $(m_s)$ as a function  of Q: (a) Effect of system size for fixed $\rho$ : (i) L = 100, (ii) L = 200, (iii) L = 400. Here $\rho = 0.5$. (b) Effect of particle density for fixed L: (i) $\rho =0.1$, (ii) $\rho = 0.3$, (iii) $ \rho = 0.5$. Here L = 400. (c) Probability of the largest size cluster $(m_s)$ as a function  of scaled variable $Q_s = Q/L^2 \rho ( 1 - \rho)$ : (i) L=100, $\rho = 0.5$, (ii) L=200, $\rho = 0.5$, (iii) L=400, $\rho = 0.5$, (iv) L=400, $\rho = 0.3$, (v) L=400, $\rho = 0.1$. Here the data points corresponds to the plots of Fig 6(a) and Fig6(b). MC simulations were performed and averaging was done over 10000 samples }
    	\label{fig6}
    \end{figure*}

  We next study the finite size effects  on the cluster size distribution. In order to investigate the finite size effects on the cluster size distribution, we perform MC simulations for relatively small system size $L$. First we analyze the evolution of the cluster size distribution for such finite size system upon variation of $Q$. In particular we observe that for a fixed system size $L$ , for sufficiently large value of $Q$,  a secondary peak starts appearing corresponding to largest size cluster,  $m_s = \rho L$  in the cluster size distribution. For a fixed $L$, Increase in $Q$ leads to increase in the height of this secondary peak which eventually starts dominating the cluster size distribution as shown in Fig.\ref{fig5}. Indeed it can be argued for any fixed  finite size system with $L$ sites, in $Q \rightarrow \infty$, the largest cluster would have probability 1. This maybe understood in following manner:  Any switching event at the end of the largest size cluster which results in decrease of the size of the cluster (due to the escape of a single particle from the cluster end) would {\it instantly} be compensated by the same particle joining the other end of the cluster, and consequently restoring the size of the largest size cluster. 
However, for a fixed Q, the single cluster of size $m_s$ always ceases to exist for a large enough system size. The appearance of this  secondary peak is simply a manifestation of  finite size  effects in contrast to genuine phase transition that is seen for many class of driven systems including Zero Range Process \cite{intro3, pt,pt1,pt2,pt3}. In Fig.\ref{fig6}(a), we plot the probability of this secondary peak corresponding to maximum size cluster, $m_s$ as a function of $Q$ for different system sizes. As expected, on increasing system size, the appearance of this peak occurs for a relatively higher value of $Q$. In Fig.\ref{fig6}(b), we show the effect of variation of particle density on  the plot of $P(m_s)$ vs $Q$. While our numerical analysis establishes that the appearance of a single  large cluster can be attributed to finite size effect  nevertheless, it is interesting to note that  it is possible to obtain a data collapse for $P(m_s)$ vs $Q$ for different L and $\rho$ when the Q axis  is rescaled to $Q/[L^{2} \rho(1 - \rho)]$. This is shown in Fig.\ref{fig6}(c).

\section{Conclusion and Outlook}
 In summary, in this paper we have studied a two species driven 
lattice gas model in which the two types of particles move
in opposite directions with rate $v$ on a 1D lattice of size $L$ lattice sites. Stochastic directional switching of particles occurs with a rate ($\alpha$) whenever the adjacent site is occupied by another particle. We study the steady state  distribution of clusters of particles as a function of the ratio $Q=v/\alpha$ of the translation and the  switching rates. We show that when the switching rate of particles is much faster than translation rate of the particles, corresponding to $Q \rightarrow 0$ limit, for $L\rightarrow \infty$, the cluster size distribution has an exponential form with a mean cluster size $\langle m \rangle = 1/(1- \rho)$. In the limit of  $L \rightarrow \infty$, for $Q >> 1$, while the cluster size distribution itself changes, with the mean cluster size  $\langle m \rangle \propto  Q^{\frac{1}{2}}$,  the distribution continues to exhibit an exponential form. In this is limit our model can be mapped to PEP discussed in Ref.\cite{soto} and an approximate analytic expression for the mean cluster size in the dense phase can be obtained which explains the $\sqrt{Q}$ dependence of the mean cluster size in this limit. 

 We also observe that for finite size lattice, when $Q >> L^{2}$, all the particles tend to stick to form  a single large cluster of size $m_s=\rho L$ and the corresponding probability distribution of the cluster size, $P(m)$,  exhibits a secondary peak at the value of  cluster size $ m_s = \rho L$. Although the formation of the single large cluster for larger values of  $Q$ is reminiscent of the clustering transition phenomena observed in a number of non-equilibrium two species models \cite{pt3,basu1}, we note that this is really a finite size effect. Indeed the large aggregate cluster vanishes in the thermodynamic limit $ L \rightarrow \infty$, for any given $Q$. Interestingly, we find that the probability of this large cluster $P(m_s)$, exhibits a distinct kind of scaling behaviour such that  in terms of scaled variable $Q_s = Q/L^2\rho(1 - \rho)$, we observe a data collapse for the plot of $P(m_s)$ vs $Q_s$. It remains an open problem to unravel the underlying physical principle which governs this finite size scaling behaviour that we have observed numerically. 
  
While the focus of our work has been restricted to analyzing the nature of distribution of clusters for the underlying lattice model, it would interesting to see what extent the generic features related to cluster and aggregate formation that we have discussed in this paper are relevant in context of driven biological processes occurring in 1D geometry such as that of migration of cells in 1D collision assays.  However it is important to recognize, that a coherent understanding for the specific active system would entail incorporating further details related to the phenomenology of such systems. For instance, for MDMK cells in such 1D collision assays,  the individual cells  tend to align them away from neighboring cells - a phenomenon referred to as Contact Inhibition Locomotion(CIL) \cite{ananyo,CIL}. Suitable modification of the our minimal model, which incorporates CIL phenomenology apart from other details of cell-cell interactions may help in understanding  the generic characteristics of collective organization and migrations of cells in such confined geometry.

\noindent
{\em Acknowledgment} Financial support is acknowledged by SM for SERB project No. EMR /2017/001335. SM also acknowledges financial support and hospitality for visit to ICTP, Trieste under the Associateship program, where part of the work was done.

\vskip 0.5cm
\noindent
All authors have equally contributed to the work.

\bibliographystyle{prsty}

\end{document}